\documentclass[aps,prb,twocolumn,groupedaddress,amsfonts,showpacs]{revtex4}
\usepackage{graphicx,color}
\usepackage{bm}

\begin{document}
\title{Spin-orbit coupling induced by a mass gradient}
\author{A. Matos-Abiague}
\affiliation{Institute for Theoretical Physics, University of
Regensburg, 93040 Regensburg, Germany}
\date{\today}

\begin{abstract}
The existence of a spin-orbit coupling (SOC) induced by the
gradient of the effective mass in low-dimensional heterostructures
is revealed. In structurally asymmetric quasi-two-dimensional
semiconductor heterostructures the presence of a mass gradient
across the interfaces results in a SOC which competes with the SOC
created by the electric field in the valence band. However, in
graded quantum wells subjected to an external electric field, the
mass-gradient induced SOC can be finite even when the electric
field in the valence band vanishes.
\end{abstract}

\pacs{71.70.Ej,73.61.Ey,73.20.Qt,73.21.Fg} \keywords{spin-orbit
coupling, semiconductor heterostructures}

\maketitle

%\section{Introduction}
%\label{intro}

Semiconductor spintronics is an emerging field based on the
controlled manipulation of the carrier spins for data processing
and device operations.\cite{Zutic2004:RMP,Fabian2007:APS} Most
proposals for spintronic devices rely on the ability of
manipulating electron spins by using the spin-orbit coupling
(SOC), which is the most fundamental spin-dependent interaction in
nonmagnetic semiconductors.\cite{Fabian2007:APS}
%The SOC is also at the core of diverse
%phenomena such as spin relaxation,\cite{...} spin Hall
%effect,\cite{...} spin Coulomb drag,\cite{...} spin-galvanic
%effect,\cite{...} spin ratchets,\cite{...} and beating of Friedel
%oscillations,\cite{...} among others. Therefore, understanding the
%nature of the different mechanisms behind the SOC is essential not
%only from the fundamental point of view but also for the
%development of spintronic devices.
%
In semiconductor heterostructures the SOC results from the lack of
inversion symmetry. The bulk inversion asymmetry (BIA) of
zinc-blende semiconductors leads to the so-called Dresselhaus
SOC,\cite{Dresselhaus1955:PR} while the structure inversion
asymmetry (SIA) of the heterostructure itself results in the
Bychkov-Rashba (BR) SOC.\cite{Bychkov1984:JPC}

In this paper I focus on the investigation of the BR-type SOC and
show that, in addition to the SOC generated by the electric field
in the valence
band,\cite{Lassnig1985:PRB,Winkler:2003,Winkler2004:PE} there is a
mass-gradient contribution to the SIA-induced SOC. In general, the
two contributions compete. However, in some specific cases the
mass-gradient induced SOC dominates.

The emergence of SOC due to the existence of a mass gradient can
be better understood by establishing an analogy between the
nonrelativistic limit of Dirac's theory and the effective-mass
Hamiltonian for conduction electrons in semiconductor
heterostructures.

For simplicity and without loss of generality, I consider the case
of a time-independent system in the presence of an electrostatic
potential $V$. Starting with the time-independent Dirac equation
the upper $\psi_u$ and lower $\psi_l$ components of the
four-component spinor $\Psi=(\psi_{u},\psi_{l})^T$ are found to be
coupled through the equations\cite{Sakurai:1967}
\begin{equation}\label{dirac1}
    (\mbox{\boldmath$\sigma$}\cdot \mathbf{p})\psi_{l}=\frac{1}{c}(\epsilon-V)\psi_{u},
\end{equation}
\begin{equation}\label{dirac2}
    (\mbox{\boldmath$\sigma$}\cdot \mathbf{p})\psi_{u}=\frac{1}{c}(\epsilon-V+2m_{0}c^2)\psi_{l},
\end{equation}
where $\epsilon$ is the particle energy (measured from the rest
energy $m_{0}c^2$), $\mbox{\boldmath$\sigma$}$ is the vector of
Pauli matrices and $\mathbf{p}$, $m_{0}$, and $c$, refer to the
momentum operator, the bare electron mass, and the velocity of
light, respectively. It follows from Eq.~(\ref{dirac2}) that
$\psi_l$ is smaller than $\psi_u$ by a factor $\sim v/c$. Thus, in
the nonrelativistic limit ($v\ll c$) the main contribution to the
four-component spinor comes from $\psi_u$. Using
Eq.~(\ref{dirac2}) one can eliminate $\psi_l$ from
Eq.~(\ref{dirac2}) and obtain, after some algebra, an equation
that involves only $\psi_u$,
\begin{equation}\label{dirac-soc}
    \left\{\mathbf{p}\cdot\left(\frac{1}{2\mu}\mathbf{p}\right)+
    \frac{\hbar}{2}\left[\mbox{\boldmath$\nabla$}\left(\frac{1}{\mu}\right)\times
    \mathbf{p}\right]\cdot\mbox{\boldmath$\sigma$}+V\right\}\psi_{u}=\epsilon \psi_{u},
\end{equation}
where I have introduced the position-dependent \emph{potential}
mass
\begin{equation}\label{p-mass}
    \mu=m_{0}\left(1+\frac{\epsilon-V}{2m_{0}c^2}\right).
\end{equation}
Note that no approximation has been made in deriving
Eq.~(\ref{dirac-soc}). The two-component spinor $\psi_u$ is,
however, not normalized. The standard procedure to overcome this
problem in the nonrelativistic limit is to introduce a new
normalized two-component spinor
$\tilde{\psi}=(1+p^{2}/8m_{0}c^{2})\psi_u$.\cite{Sakurai:1967} In
doing this, however, some approximations has to be made and the
Hamiltonian for the normalized spinor $\tilde{\psi}$ acquires
additional terms.\cite{Sakurai:1967} However, these additional
terms are irrelevant for our discussion here and will be omitted
in our analysis.

The Hamiltonian in Eq.~(\ref{dirac-soc}) is Hermitian and
resembles the effective-mass Hamiltonian describing the motion of
electrons in a solid with position-dependent effective mass.
Interestingly, the spin-orbit coupling seems to originate from the
gradient of the potential mass $\mu$. In the Dirac theory the
energy gap, $E_{0}=2m_{0}c^2$, which separates the energy spectra
of the free particles and antiparticles is a position-independent
constant. Therefore, the only position dependence in $\mu$ which
can lead to a finite SOC has to come from the electrostatic
potential $V$. Thus, in the Dirac theory, the SOC emerges purely
from the electric field
$\mathbf{E}=(-1/e)\mbox{\boldmath$\nabla$}V$ (here $e$ denotes the
electron charge). In the nonrelativistic approximation the
dominant energy is the vacuum gap $E_0$ and the inverse potential
mass can be approximated as [see Eq.~(\ref{p-mass})]
\begin{equation}\label{inv-mass}
    \frac{1}{\mu}\approx
    \frac{1}{m_0}+\frac{V-\epsilon}{2m_{0}^2c^{2}}.
\end{equation}
As a result the SOC reduces to the well-known
form\cite{Sakurai:1967}
\begin{equation}\label{soc-D}
    H_{\rm{so}}=\frac{\hbar}{2}\left[\mbox{\boldmath$\nabla$}\left(\frac{1}{\mu}\right)\times
    \mathbf{p}\right]\cdot\mbox{\boldmath$\sigma$}\approx \frac{\hbar c^2}{E_{0}^2}\left(\mbox{\boldmath$\nabla$}V\times
    \mathbf{p}\right)\cdot\mbox{\boldmath$\sigma$}.
\end{equation}

In the case of semiconductors the analogue to the potential mass
$\mu$ is the effective mass $m^{\ast}$, while the equivalent to
the vacuum gap $E_0$ is the energy gap $E_g$ separating the energy
spectrum of electrons in the conduction band from the hole
spectrum in the valence band. In contrast to the vacuum, where
$E_0$ is a constant, in semiconductor heterostructures the energy
gap $E_g$ becomes position dependent. Therefore, the position
dependence of the effective mass may originate from both the
electrostatic potential $V$ and the band gap $E_g$. As a result,
in addition to the conventional SOC produced purely by the
electric field a finite SOC contribution induced by the position
dependence of the effective mass emerges.

To investigate in more details the mass gradient induced SOC, I
consider a semiconductor heterostructure grown in the $z$
direction. In such a case the mass gradient induced SOC can be
related to the well-known Bychkov-Rashba (BR) SOC observed in
quasi two-dimensional systems with structure inversion asymmetry
(SIA).\cite{Bychkov1984:JPC}

\begin{figure}
\includegraphics*[width=6cm]{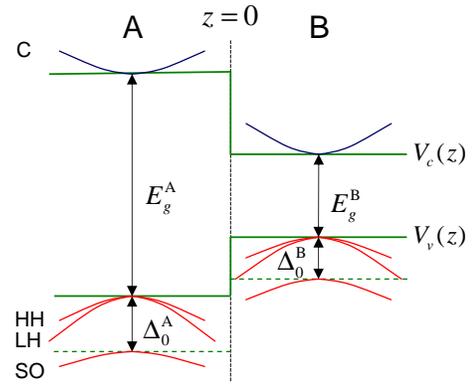}
\caption{(color online). Schematic of the band-edge profile of an
A/B semiconductor interface located at $z=0$. The conduction,
heavy hole, light hole, and split-off bands are labelled as C, HH,
LH, and SO, respectively.} \label{F1}
\end{figure}

The effective Hamiltonian describing the motion of the conduction
band electrons in the heterostructure can be obtained by using the
envelope function approximation. I consider the ($8\times 8$) Kane
Hamiltonian\cite{Kane1980:LNP,Winkler:2003,Fabian2007:APS} which
accounts for the $\Gamma_{6c}$, $\Gamma_{8v}$, and $\Gamma_{7v}$
bands [see bands C, HH and LL, and SO bands, respectively, in
Fig.~\ref{F1}]. The conduction and valence band states can be
decoupled by using the folding-down (L\"{o}wdin)
technique.\cite{Lowdin1951:JCP,Winkler:2003,Fabian2007:APS}
Neglecting the non-parabolicity effects, the effective Hamiltonian
for the conduction electrons is found to
be\cite{Lassnig1985:PRB,Silva1997:PRB,Fabian2007:APS}
\begin{equation}\label{h-eff}
    H_{\rm{eff}}=\frac{p_{\parallel}^{2}}{2m^{\ast}(z)}-\frac{\hbar^{2}}{2}
    \frac{d}{dz}\left[\frac{1}{m^{\ast}(z)}\frac{d}{dz}\right]+V_{c}(z)+H_{\rm{so}},
\end{equation}
where
%\begin{eqnarray}\label{m-eff}
\begin{equation}\label{m-eff}
    \frac{1}{m^{\ast}(z)}=\frac{1}{m_{0}}-\frac{2\tilde{P}^2}{3m_{0}^2}
    \left[\frac{2}{V_{v}(z)}+\frac{1}{V_{v}(z)-\Delta_{0}(z)}\right]
    %\nonumber
    %\\
    %&+&\frac{1}{m_{0}}
%\end{eqnarray}
\end{equation} is the $z$-dependent, inverse effective mass for the
conduction band electrons and
\begin{equation}\label{hso-eff}
    H_{\rm{so}}=\frac{\alpha(z)}{\hbar}(p_{y}\sigma_{x}-p_{x}\sigma_{y})
\end{equation}
with
\begin{equation}\label{alpha}
\alpha(z)=\frac{\hbar^{2}}{3m_{0}^{2}}\frac{d}{dz}
\left[\frac{\tilde{P}^{2}}{V_{v}(z)}-\frac{\tilde{P}^{2}}{V_{v}(z)-\Delta_{0}(z)}\right]
\end{equation}
is the SIA-induced SOC. In the equations above $p_{x}$ and $p_y$
are the components of the in-plane momentum
$\mathbf{p}_{\parallel}$ and $\tilde{P}=\langle S|p_{x}|P\rangle$
represents the non-vanishing momentum matrix elements involving
the s-like band edge Bloch state ($|S\rangle$) of the conduction
band and the p-like hole states
($|P\rangle=|X\rangle,|Y\rangle,|Z\rangle$). The correction to the
effective mass due to the interaction with remote bands can be
included by using perturbation
theory.\cite{Vurgaftman2001:JAP,Winkler:2003} As shown in
Fig.~\ref{F1}, $\Delta_{0}(z)$ refers to the spin-orbit splitting
energy, while $V_{c}(z)$ and $V_{v}(z)$ are the potential profiles
of the conduction and valence band edges, respectively.

For the case of a quantum well (QW) grown along the $z$-direction
one can obtain an effective Hamiltonian describing the in-plane
motion of a tow-dimensional electron gas by averaging
Eq.~(\ref{h-eff}) with the spin-independent $z$-component, $f(z)$,
of the wave function. This results in the so-called BR SOC with
$\alpha_{_{\rm BR}}=\langle \alpha(z)\rangle_c=\int \alpha(z)
|f(z)|^{2}dz$ as the BR SOC strength.

At first glance it seems that the BR spin splitting should be
proportional to the electric field which brakes the spatial
inversion symmetry. However, the fact that for quasi 2D systems
with position-independent effective mass the electric field along
the direction of confinement must average to zero\cite{Darr:1976}
(this follows from Ehrenfest's theorem which states that the
average force on a bound state vanishes\cite{Sakurai:1994})
generated intensive discussions about the nature of the electric
field causing the BR
SOC.\cite{Ohkawa1974:JPSJ,Darr:1976,Lassnig1985:PRB,Malcher1986:SM}
It was latter found that for a quantum well growth along the $z$
direction, with $z$-dependent effective mass, the average electric
field ${\rm E}_{c}\sim \langle
p_{z}[(\partial_{z}m^{-1})p_{z}]\rangle_c$ may not
vanish.\cite{Zawadzki2001:PRB} However, the estimated value for
this field was found to be too
small\cite{Malcher1986:SM,Zawadzki2001:PRB} as to explain the
experimentally observed spin-splitting due to the SIA SOC.
Actually, the SOC induced by ${\rm E}_{c}$ represents a high-order
correction which is not even present in the effective Hamiltonian
obtained with the standard $(8\times 8)$ Kane approximation.

In an attempt to clarify the origin of the BR SOC,
Lassnig\cite{Lassnig1985:PRB} showed that the BR spin splitting in
the conduction band is related to the electric field in the
valence band [$\mathbf{E}_v=(-1/e)\mbox{\boldmath$\nabla$}V_v$]
whose average (over the conduction states) does not necessarily
vanish (note that in this case Ehrenfest's theorem does not
apply\cite{Winkler:2003,Winkler2004:PE}). Below I show that in
addition to the SOC induced purely by $\mathbf{E}_v$ a
contribution originating from the existence of a mass gradient
appears. I remark, however, that the mass-gradient contribution
discussed here is much larger than the one generated by the
mass-gradient induced electric field ${\rm E}_c$ (see discussion
in the previous paragraph) and appears readily in the effective
Hamiltonian resulting from the $(8\times 8)$ Kane model. In fact,
the BR SOC can be reinterpreted as resulting from the competition
between the here proposed mass-gradient SOC and the SOC induced
purely by $\mathbf{E}_{v}$. Interestingly, the mass gradient
contribution (and therefore the BR SOC) can be finite even when
the electric field in the valence band vanishes (i.e,
$\mathbf{E}_v=0$).

\begin{figure}
\includegraphics*[width=5cm]{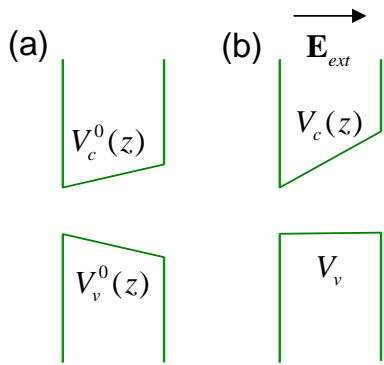}
\caption{(color online). Schematics of the conduction- and
valence-band potentials in a linearly graded quantum well with
infinite barriers, in the absence (a) and presence (b) of an
external electric field $\mathbf{E}_{ext}$. The external electric
field compensates the electric field in the valence band in such a
way that the total electric field in the valence band vanishes
(i.e., $\mbox{\boldmath$\nabla$}V_{v}^{f}=0$).} \label{F2}
\end{figure}

Combining Eqs.~(\ref{m-eff}) and (\ref{alpha}) one can rewrite the
SOC parameter as
\begin{equation}\label{alpha-new}
    \alpha(z)=\frac{\hbar^{2}}{2}\frac{d}{dz}\left[\frac{1}{m^{\ast}(z)}\right]+
    \frac{\hbar^{2}}{m_{0}^{2}}\frac{d}{dz}\left[\frac{\tilde{P}^{2}}{V_{v}(z)}\right].
\end{equation}
When modelling semiconductor heterostructures, the momentum matrix
element, $\tilde{P}$, is commonly considered to be position
independent.\cite{Winkler:2003} Within this approximation and
taking into account that for most semiconductors $V_{v}$ is of the
same order of the band gap (i.e., $|V_{v}-E_g|\ll E_{g}$, one can
approximate Eq.~(\ref{alpha-new}) as
\begin{equation}\label{alpha-new2}
    \alpha(z)=\frac{\hbar^{2}}{2}\frac{d}{dz}\left[\frac{1}{m^{\ast}(z)}\right]-
    \frac{\hbar^{2}(\tilde{P}/m_{0})^{2}}{E_{g}^{2}}\frac{d}{dz}V_{v}(z).
\end{equation}
Thus, one can rewrite
% Eq.~(\ref{alpha-new}) as
%\begin{equation}\label{alpha-tot}
%    \alpha(z)=\alpha_{m}(z)+\alpha_{v}(z),
%\end{equation}
%where
%\begin{equation}\label{alpha-mass}
%    \alpha_{m}(z)=\frac{\hbar^{2}}{2}\frac{d}{dz}\left[\frac{1}{m^{\ast}(z)}\right],
%\end{equation}
%and
%\begin{equation}\label{alpha-pot}
%    \alpha_{v}(z)=-
%    \frac{\hbar^{2}(\tilde{P}/m_{0})^{2}}{E_{g}^{2}}\frac{d}{dz}V_{v}(z).
%\end{equation}
%Similarly,
the SOC in Eq.~(\ref{hso-eff}) as
$H_{\rm{so}}=H_{\rm{so}}^{m}+H_{\rm{so}}^{v}$, where
\begin{equation}\label{hso-mass}
    H_{\rm{so}}^{m}=\frac{\hbar}{2}\left[\mbox{\boldmath$\nabla$}\left(\frac{1}{m^{\ast}}\right)\times
    \mathbf{p}\right]\cdot\mbox{\boldmath$\sigma$}
\end{equation}
is the mass gradient induced SOC and
\begin{equation}\label{hso-pot}
    H_{\rm{so}}^{v}=-\frac{\hbar (\tilde{P}/m_{0})^{2}}{E_{g}^2}\left(\mbox{\boldmath$\nabla$}V_{v}\times
    \mathbf{p}\right)\cdot\mbox{\boldmath$\sigma$}
\end{equation}
is the SOC created by the electric field in the valence band.

Equation (\ref{hso-pot}) has the same structure as
Eq.~(\ref{soc-D}),\cite{note} which confirms that, indeed,
$H_{\rm{so}}^{v}$ corresponds to the \emph{standard} SOC
generated, purely, by an electric field; in this case by the
valence-band electric field $\mathbf{E}_{v}=(-1/e)
\mbox{\boldmath$\nabla$}V_{v}$. Consequently, $H_{\rm{so}}^{v}$
vanishes when $\mathbf{E}_{v}=0$. On the contrary, it is clear
from Eqs.~(\ref{m-eff}) and (\ref{hso-mass}) that even in the case
of vanishing $\mathbf{E}_v$, the mass gradient induced SOC,
$H_{\rm{so}}^{m}$, remains, in general, finite.

For an estimation of the strengths of the mass gradient and
valence-band electric field induced SOCs, I consider the case of
an A/B abrupt interface between two III-V semiconductors [see
Fig.~\ref{F1}]. In such a case the band parameters, and therefore
$m^{\ast}$ and $V_{v}$, are step-like functions of $z$ and the
mass-gradient and valence-band electric field induced SOCs in
Eq.~(\ref{alpha-new}) reduce to
\begin{equation}\label{hso-inter}
    H_{\rm{so}}^{m,v}=\frac{\alpha_{m,v}}{\hbar}\delta(z)(p_{y}\sigma_{x}-p_{x}\sigma_{y})
\end{equation}
with
\begin{equation}\label{alpha-mass}
    \alpha_{m}=\frac{\hbar}{2}\left(\frac{1}{m_{\rm{B}}^{\ast}}-\frac{1}{m_{\rm{A}}^{\ast}}\right),
\end{equation}
and
\begin{equation}\label{alpha-pot}
    \alpha_{v}=\frac{\hbar^{2}}{m_{0}^{2}}\left(\frac{\tilde{P}_{\rm A}^{2}}{E_{g}^{\rm{A}}}-
    \frac{\tilde{P}_{\rm B}^{2}}{E_{g}^{\rm{B}}}\right),
\end{equation}
respectively. Here the values of the parameters in regions A and B
are indicated with the respective labels. Note also, that for a
better accuracy, the step-like position dependence of the momentum
matrix element has also been considered in Eq.~(\ref{alpha-pot}).

The calculated values of $\alpha_m$, $\alpha_v$, and the total
interface SOC strength, $\alpha_{\rm{int}}=\alpha_{m}+\alpha_{v}$,
for different interfaces are listed in Table \ref{tab1}. For all
the considered interfaces, $\alpha_m$ and $\alpha_v$ are of the
same order but with opposite signs. Thus, the competition between
the mass-gradient and valence-band electric field induced SOC
contributions results in the decrease of the total interface SOC.
%The results for $\alpha_{\rm{int}}$ are in good agreement with the
%corresponding values derived in Ref.~\onlinecite{Fabian2007:APS}.

\begin{table*}
\caption{Interface spin-orbit coupling parameters (in
eV$\textrm{\AA}^{2}$) for A/B abrupt interfaces composed of
arsenides. $\alpha_m$ and $\alpha_v$ correspond to the
contributions due to the mass gradient and valence-band electric
field, respectively, while $\alpha_{\rm{int}}$ is the total
interface SOC strength.} \label{tab1}
\begin{ruledtabular}
\begin{tabular}{ccccc}
  %\hline
  A & B & $\alpha_m$ & $\alpha_v$ & $\alpha_{\rm{int}}$ \\
  \hline
  AlAs & InAs & 138.93 & -169.02 & -30.09 \\
  %
  %\hline
  GaAs & AlAs & -35.88 & 39.07 & 3.19 \\
  %
  %\hline
  InAs & GaAs & -103.05 & 129.95 & 26.9 \\
\end{tabular}
\end{ruledtabular}
\end{table*}

In systems in which $\alpha_m$ and $\alpha_v$ are of the same
order (as the ones considered above) the mass-gradient
contribution to the SOC is masked by the SOC induced by the
valence-band electric field. Therefore the experimental
measurement of $\alpha_m$ alone may be difficult in such systems.
To overcome this problem, I propose to measure $\alpha_m$ in a
graded, semiconductor quantum well subjected to an external
electric field.

For illustration, I consider a ${\rm Ga}_{1-x}{\rm Al}_{x}{\rm
As}$-based quantum well with high potential barriers, so that
interface effects play a little role and can be neglected. In a
linearly graded quantum well (i.e, with Al concentration varying
linearly with the position) the energy gap together with the band
parameters become position dependent. For small grading, the band
parameters interpolate linearly with the Al concentration.
Consequently, both the potential profile of the conduction
($V_{c}^{0}$) and valence ($V_{v}^{0}$) band edges change linearly
with $z$ [see Fig.~\ref{F2}(a)]. In the presence of a constant
external electric field, $\mathbf{E}_{ext}$, oriented along the
growth direction the band edge profiles are modified as
$V_{c}=V_{c}^{0}(z)-e{\rm E}_{ext}z$ and $V_{v}=V_{v}^{0}(z)-e{\rm
E}_{ext}z$. Since $V_{v}^{0}(z)$ is a linear function of $z$ one
can find a \emph{target} electric field for which $V_{v}$ becomes
position independent and the total electric field in the valence
bands vanishes [see Fig.~\ref{F2}(b)]. Therefore, for such a
target external field, $\alpha_v \approx 0$, while $\alpha_m$
remains finite. Under this condition, the SOC is determined solely
by the mass-gradient contribution.

I now estimate the values of $\alpha_m$ and the target external
field ${\rm E}_{\rm target}$ for a quantum well with Al
concentration varying linearly from $x_{\rm min}=0$ to $x_{\rm
max}=0.1$, i.e.,
\begin{equation}\label{x-to-z}
    x(z)=x_{\rm max}z/d.
\end{equation}
Here $d=200\textrm{ \AA}$ is the well width. The composition
dependence of the band parameters is evaluated according to the
interpolation scheme developed in
Ref.~\onlinecite{Vurgaftman2001:JAP}. I then expand
Eq.~(\ref{m-eff}) up to the linear order in $x$ [which for the
small concentrations considered here ($x\leq 0.1$) suffices] and
obtain the position dependence of the inverse effective mass by
using (\ref{x-to-z}). For the target external field ${\rm E}_{\rm
target}\approx 30 \textrm{ kV/cm}$ the valence-band electric field
vanishes (i.e., $\alpha_{v}=0$), while the strength of the
mass-gradient induced SOC is found to be $\alpha_{m}\approx -8.74
\textrm{ meV \AA}$. Apart from the sign, this value is of the same
order but still larger than the SIA SOC parameters experimentally
measured in GaAs-AlGaAs asymmetric quantum
wells.\cite{Jusserand1995:PRB,Miller2003:PRL}

Beyond the case of semiconductor heterostructures, the
mass-gradient induced SOC is expected to be relevant in
metal/semiconductor interfaces across which the values of the
effective mass have large and abrupt changes.
%The mass-gradient
%induced SOC may also be dominant in interfaces between inverted
%gap semiconductors (e.g.,), where the sign of the effective mass
%changes across the interface.\cite{...} In fact, large SOC fields
%have already been measured in ...\cite{...}

In summary, I have revealed the existence of a mass-gradient
contribution to the SOC in systems with structure inversion
asymmetry. I have shown that for some semiconductor
heterostructures, the mass-gradient contribution is of the same
order as the SOC generated by the electric field in the valence
band. However, in the particular case of a linearly graded
semiconductor quantum well subjected to a conveniently designed
external field, the electric field in the valence band vanishes
and the remaining SOC is purely induced by the mass gradient.

\acknowledgments{I am grateful to J. Fabian for fruitful
discussions. This work was supported by the Deutsche
Forschungsgemeinschaft via SFB 689.}

\bibliographystyle{apsprb}

%\bibliography{references}

%\begin{references}
%\begin{thebibliography}{00}

\end{document}